\documentclass[aps,prl,Letter,twocolumn,superscriptaddress,showpacs,amsmath,amssymb]{revtex4}

\usepackage{graphicx}
\usepackage[normalem]{ulem} \usepackage{color}
\definecolor{DarkBlue}{rgb}{0.1,0.1,0.5} \definecolor{Red}{rgb}{0.9,0.0,0.1}
\definecolor{Green}{rgb}{0.0,0.99,0.0}

\begin{document}

\title{The Role of Lattice Coupling in Establishing Electronic and Magnetic
Properties in Quasi-One-Dimensional Cuprates}

\author{W. S. Lee} \email{leews@stanford.edu} \affiliation {SIMES, SLAC
National Accelerator Laboratory, Menlo Park, CA 94025, USA}

\author{S. Johnston} \affiliation {SIMES, SLAC National Accelerator Laboratory,
Menlo Park, CA 94025, USA} \affiliation{IFW Dresden P.O. Box 27 01 16, D-01171
Dresden, Germany}

\author{B. Moritz} \affiliation{SIMES, SLAC National Accelerator Laboratory,
Menlo Park, CA 94025, USA} \affiliation{Department of Physics and Astrophysics,
University of North Dakota, Grand Forks, ND 58202, USA} \affiliation{Department
of Physics, Northern Illinois University, DeKalb, IL 60115, USA}

\author{J. Lee} \affiliation{Department of Applied Physics, Stanford
University, Stanford, CA 94305, USA}

\author{M. Yi} \affiliation{Department of Applied Physics, Stanford University,
Stanford, CA 94305, USA}

\author{K. J. Zhou} \affiliation{Paul Scherrer Institut, Swiss Light Source,
CH-5232 Villigen PSI, Switzerland}

\author{T. Schmitt} \affiliation{Paul Scherrer Institut, Swiss Light Source,
CH-5232 Villigen PSI, Switzerland}

\author{L. Patthey} \affiliation{Paul Scherrer Institut, Swiss Light Source,
CH-5232 Villigen PSI, Switzerland}

\author{ V. Strocov} \affiliation{Paul Scherrer Institut, Swiss Light Source,
CH-5232 Villigen PSI, Switzerland}

\author{K. Kudo} \affiliation{Department of Applied Physics, Tohoku University,
Japan}

\author{Y. Koike} \affiliation{Department of Applied Physics, Tohoku
University, Japan}

\author{J. van den Brink} \affiliation{IFW Dresden P.O. Box 27 01 16, D-01171
Dresden, Germany}

\author{T. P. Devereaux} \affiliation{SIMES, SLAC National Accelerator
Laboratory, Menlo Park, CA 94025, USA}

\author{Z. X. Shen} \affiliation{SIMES, SLAC National Accelerator Laboratory,
Menlo Park, CA 94025, USA}


\date{\today}

\begin{abstract} High resolution resonant inelastic x-ray scattering has been
    performed to reveal the role of lattice-coupling in a family of quasi-1D
    insulating cuprates, Ca$_{2+5x}$Y$_{2-5x}$Cu$_5$O$_{10}$. Site-dependent
    low energy excitations arising from progressive emissions of a 70 meV
    lattice vibrational mode are resolved for the first time, providing a
    direct measurement of electron-lattice coupling strength. We show that such
    electron-lattice coupling causes doping-dependent distortions of the
    Cu-O-Cu bond angle, which sets the intra-chain spin exchange interactions.
    Our results indicate that the lattice degrees of freedom are fully
    integrated into the electronic behavior in low dimensional systems.
\end{abstract}

\pacs{Valid PACS appear here}
\maketitle

\begin{figure*} [t] \includegraphics [clip, width=6.5 in]{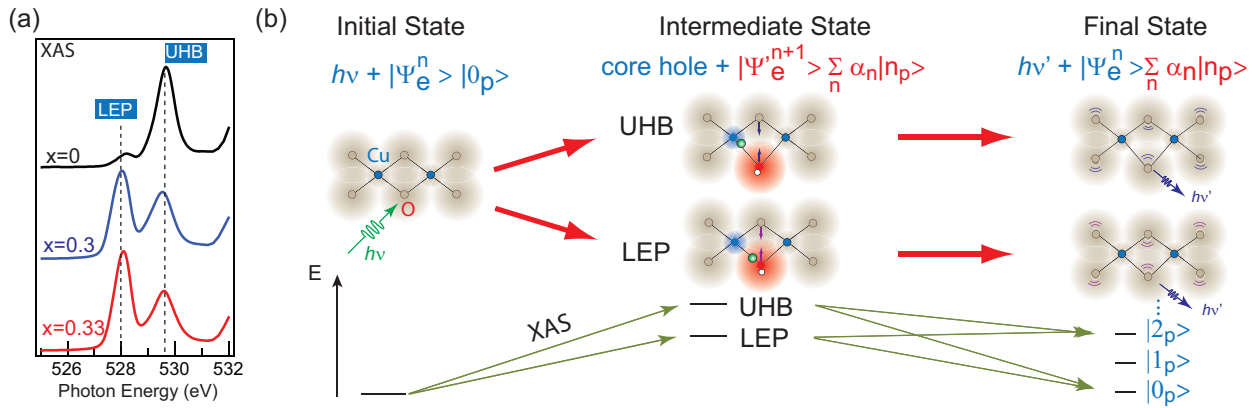}
    \caption{\label{Fig1:Illustration} (color online) (a) x-ray absorption
        spectrum near the oxygen \emph{K}-edge for
        Ca$_{2+5x}$Y$_{2-5x}$Cu$_5$O$_{10}$ compounds of three different doping
        levels.  The ``UHB'' and ``LEP" denotes the absorption peaks at 529.5
        and 528 eV, respectively. The average number of hole per Cu ion is
        denoted as x.  (b)$|\Psi_{e}^n>$ and $|\Psi_{e}^{n+1}>$ denotes the
        electronic ground state and the excited electronic intermediate state
        with an additional electron excited from the oxygen $1s$ core level.
        $|n_p>$ ($n$ = 0, 1, 2,...) denotes the number of phonon quanta excited
        in the lattice degree of freedom with a wavefunction superposition
        coefficient $\alpha$. For purposes of illustration, the wavefunction is
        expressed as product states of the diagonalized electron and lattice
        Hamiltonians. Here, the lattice vacuum state $|0_p>$ is referenced to
        the initial equilibrium phonon occupation and $|n_p>$ describes
        excitations from this state. In the RIXS intermediate state, an
        electron (green ball) is excited, leaving a core hole (white ball) in
        the O $1s$ level. The excited electron can be closer to either Cu or O
        site by tuning the incident photon energy to match the UHB or LEP
        resonance, respectively. The change of the local charge density on Cu
        (blue shade) and O (red shade) sites induce local relaxation of the
        distorted lattice, creating a distribution of $|n_p>$. In the final
    state, the excited electron can recombine with the core hole, leaving the
lattice in excited states. These final states have non-zero overlap with the
intermediate state, yielding harmonic phonon excitations in RIXS spectrum.}
\end{figure*}

\begin{figure*}[t] \includegraphics [clip, width=6.5 in]{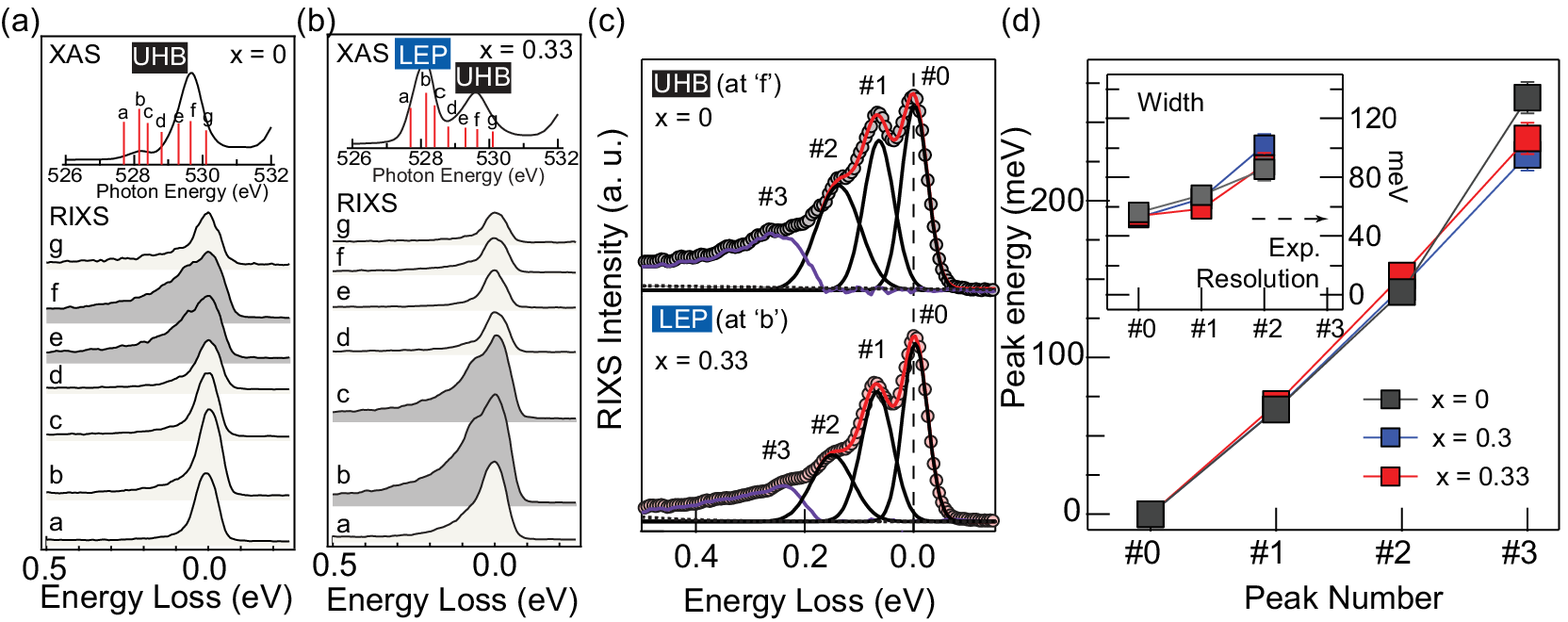}
    \caption{\label{Fig2:RIXS_data} (color online) The RIXS spectrum for (a) x
        = 0 and (b) x = 0.33 samples taken at several incident photon energies,
        as indicated by the red bars in the XAS plots. The length of these red
        vertical bars reflects the relative elastic peak intensity of the
        corresponding RIXS spectrum. The darker shaded areas highlight the
        spectra with prominent additional spectral weight near the elastic
        peak. (c) High resolution data taken at UHB resonance for the x = 0
        sample (upper) and at LEP resonance for the x = 0.33 sample (lower).
        Multi-peak features are annotated with number identifiers. The black
        curves are the Gaussian functions used for the fit (red curves). Purple
        curves are the remaining spectrum after subtracting the fit. Background
        is plotted as black dotted lines. (d) A summary plot of the fitted peak
    positions and widths (inset) of the RIXS spectra taken at UHB (x = 0) and
LEP (x = 0.3 and 0.33) resonances. All data were taken at 30 K.  }
\end{figure*}

Electron-lattice coupling is an important mechanism in solids that determines
the ground state properties via renormalizing the mass of charge carriers
\cite{Ashcroft76}, and, in some cases, induces novel symmetry-broken states
like superconductivity \cite{Tinkham96} and charge density waves
\cite{Gruner94}. While research on the electron-lattice interaction has a long
history in condensed matter physics, it continues to be an important topic in
establishing new paradigms for understanding the properties of strongly
correlated materials in which charges become more localized and screening
effects are weak \cite{Devereaux91}. To date the electron-
lattice coupling in two dimensional correlated materials such as cuprates
\cite{Lanzara01,Shen04,Lee07, Meevasana06,Johnston12} and manganites
\cite{Mannella05, Vasiliu99, Millis95} has drawn much attention in the field.
In one dimensional (1D) cuprate systems, although unusual spin and charge dynamics
\cite{Muzuno98, Voit95, Leib68, Kim06, Schlappa12} have been revealed,
studies of the role of electron-phonon coupling are still lacking.  Such studies
are important since poor screening effects in 1D systems \cite{Brink00}
should in principle cause strong electronic coupling to the lattice.

To address this issue, we study a family of quasi-1D cuprates
Ca$_{2+5x}$Y$_{2-5x}$Cu$_5$O$_{10}$, in which CuO$_2$ plaquettes arrange in a
chain-like structure by sharing their edges with neighboring CuO$_2$
plaquettes. This system is important as it is the only quasi-1D cuprate that
can be doped over a wide range of hole concentrations, providing a unique
opportunity to study doping induced phenomena \cite{hayashi98, Kudo05}. By
increasing carrier concentration, the spin ground state evolves from A-type
antiferromagnetic order (intra-chain ferromagnetic order with inter-chain
antiferromagnetic alignment), to spin glass, followed by spin gap, and
eventually to a spin disordered phase \cite{Kudo05, Matsuda05}. Curiously, the
structural Cu-O-Cu bond angle, which determines to the size and sign of the
spin super-exchange interaction according to Goodenough- Kanamori-Anderson
theory \cite{Muzuno98, Goodenough55, Kanamori59, Anderson63}, is also doping
dependent \cite{Kudo05}. However, the important mechanism behind the doping
induced Cu-O-Cu angle change remains unclear. Using high resolution resonant
inelastic x-ray scattering (RIXS), we find a 70 meV phonon strongly coupled to
the electronic state. As we will demonstrate, the doping dependent Cu-O-Cu
angle is driven by electron-lattice coupling due to the ineffectiveness of
screening in 1D. Further, the phonon energy is found to soften when cooling
across the antiferromagnetic phase transition in the undoped compounds.
These observations demonstrate that the lattice degrees of freedom are fully integrated
into the electronic and magnetic properties of low-dimensional correlated
materials.

Single crystals of Ca$_{2+5x}$Y$_{2-5x}$Cu$_5$O$_{10}$ used in the present
study were grown by traveling-solvent floating-zone (TSFZ) methods. Samples
with three different doping levels were selected for measurements, nominally x
= 0, 0.3, and 0.33. The RIXS measurements were performed at the ADRESS
beamline, Swiss Light Source, with the energy resolution set at either 50 meV
or 80 meV. The sample surface was the (0, 1, 0) plane, coincident with the
chain plane, prepared by either polishing (x = 0) or cleaving (x = 0.3 and
0.33). The scattering angle was set at 90 degree to minimize the elastic peak
intensity. As sketched in Fig. S1(c) in the Supplementary Material, the
scattering plane was the a-b plane with a grazing angle of 20 degree to the
sample surface. The polarization of the incident photon is perpendicular to the
scattering plane ($\sigma$-polarization),while the recorded spectrum includes
all the polarizations of the scattered photons.

It is informative to first introduce the doping evolution of the electronic
wave function, revealed by the x-ray absorption spectrum (XAS) near the O K-edge
($1s$-$2p$ transition). As shown in Fig. \ref{Fig1:Illustration}(a), the XAS of
the undoped compound exhibits a single dominant absorption peak (529.5 eV),
which is associated with the upper Hubbard band (UHB). Upon hole-doping, the
UHB peak decreases, indicating a reduction of the weight of the UHB component
\cite{Matsuda01}. In addition, a lower energy peak (LEP) emerges, which arises
from the spectral weight transfer to wavefunction components that are directly
related to the doped holes \cite{Okada03}. Approximately, in real space, the
XAS at the UHB resonance produces a final state with one additional valence
electron near the copper site, while the XAS at the LEP resonance puts the
additional valence electron near the oxygen site, as sketched in Fig.
\ref{Fig1:Illustration}(b).

To investigate the effects of lattice coupling on the electronic states, we use
resonant inelastic x-ray scattering (RIXS) \cite{Ament11,Ament11_2}. The
underlying principle is illustrated in Fig. \ref{Fig1:Illustration}(b). In the
O $K$-edge RIXS process, a $1s$ core electron will first be excited into an XAS
final state, either the UHB or LEP resonance.  This causes a local change in
the charge density which locally distorts the lattice. Later, the electron
de-excites, filling the O $1s$ core hole, leaving the lattice in an excited
state. The RIXS spectrum records the overlaps of the initial, intermediate, and
final states as a function of energy loss (the energy difference between the
incident and emitted photons), and includes the possibility of producing
multiple peaks in the spectrum with an energy separation corresponding to the
energy of the quanta of the lattice vibrations (i.e. phonons) in a fashion
analogous to a generalized Frank Condon picture. The spectral weight of phonon
excitations in the RIXS spectrum is directly dependent on the electron-phonon
(e-ph) coupling strength. Furthermore, the e-ph coupling strength at Cu and O
sites can be determined by tuning the incident photon energy to the UHB and LEP
resonances, respectively. This site dependency stems from differences in the
electronic character of the XAS final states.

Figure \ref{Fig2:RIXS_data}(a) displays RIXS spectra for the undoped parent
compound. As the incident photon energy is tuned to near the UHB resonance (`e'
and `f'), the spectra near the elastic scattering peak, develops an unusual
asymmetric broadening, indicating the excitations of low energy modes.  Upon
doping, as shown in Fig. \ref{Fig2:RIXS_data}(b), an asymmetric spectrum near
the elastic peak is also observed at the UHB resonance, however it is now most
pronounced at the LEP resonance. The shift of resonance behavior is associated
with the doping evolution of the XAS spectral weight due to changes of the
wavefunction character \cite{Okada03}, suggesting that the origin of the
low-energy modes is coupled directly with the electronic wavefunction.

Higher resolution RIXS spectra (Fig. \ref{Fig2:RIXS_data}(c)) provide further
information about the origin of these low-energy excitations. At both UHB and
LEP resonances, we resolved multiple peaks in the spectrum, which lie in
harmonic order of a single energy scale of ~ 70 meV (Fig.
\ref{Fig2:RIXS_data}d). We note that since the spin super-exchange energy
(10-14 meV) \cite{Matsuda01, Kuzian12} is much smaller than the extracted mode
energy, the observed excitations cannot be attributed to spin excitations.
Rather they are most likely due to an oxygen bond-stretching vibrational mode,
similar to those commonly found in 2D cuprates at a similar energy scale
\cite{Lanzara01}. In addition, the mode energy extracted from the RIXS spectra
taken at the UHB and LEP are identical, indicating that the observed phonon
excitations share the same origin. These observations demonstrate that the
electronic states in the CYCO system are coupled strongly with this 70 meV
phonon mode.

\begin{figure*} [t] \includegraphics [clip, width=6.5 in]{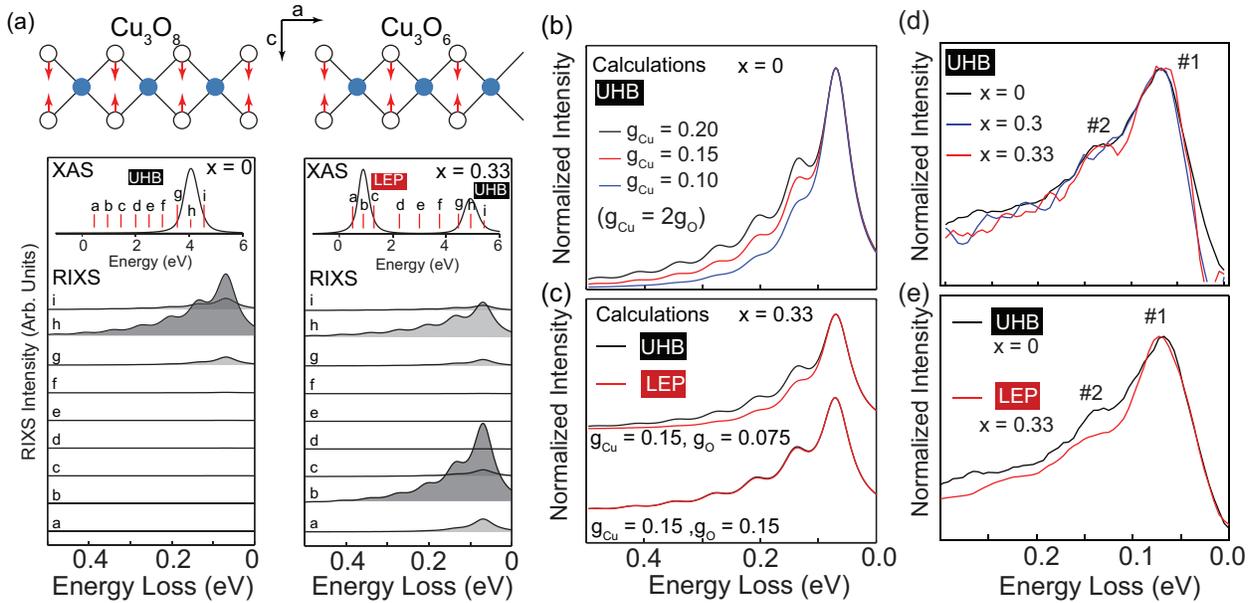}
    \caption{\label{Fig3:Theory} (color online) (a) The calculated RIXS
        intensities for the x = 0 (left) and x = 0.33 (right) hole-doped
        systems, respectively, with g = 0.15 eV. The incident photon energies
        are indicated in the calculated XAS spectra and the elastic line has
        been removed by excluding the ground state from the final state
        summation. The calculations were performed on Cu3O8 andCu3O6 clusters
        (open and periodic boundary conditions) for the x = 0 and x = 0.33
        cases, respectively. The arrows indicate the displacement pattern of
        the phonon eigenvector involving c-axis motion of the O atoms.(b) A
        comparison of the normalized phonon excitations at the UHB resonance
        for the x = 0 case and different coupling strengths. (c) The normalized
        phonon excitation spectrum at the LEP and UHB resonance in a doped
        cluster (x = 0.33). Spectra of two different values of gO were
        calculated to demonstrate the site dependence of the RIXS process at
        the LEP and UHB resonances. (d) Experimental data of harmonic phonon
        excitations (elastic peak and background subtracted) normalized to the
        first phonon peak intensity at the UHB resonance for three different
        doping levels. (e) Experimental data of the normalized harmonic phonon
    excitations for x = 0 and x = 0.33 taken at the UHB and LEP resonance,
respectively.  } \end{figure*}

To gain further insight into harmonic progression of phonon excitations, we use
exact diagonalization to calculate RIXS for CuO$_2$ clusters coupled to a
subset of optical oxygen vibrations. In this model, we include coupling to a
vibrational mode whose eigenvector is sketched in Fig. \ref{Fig3:Theory}(a)
with coupling strengths of $g_{Cu}$ and $g_O$ on Cu and O sites, respectively.
(Supplementary Material) This mode derives its coupling from the strong
electrostatic modification it produces to the electronic states of the
edge-shared CuO2 chains (Supplementary Material). Our calculations (Fig.
\ref{Fig3:Theory}(a)) capture the essential physics related to the observed
excitations, reproducing the XAS spectrum, the phonon excitations, and their
shift from the UHB to LEP resonance in the doped cluster. Importantly, we show
that their spectral weight is extremely sensitive to the e-ph coupling
strength: the stronger the coupling, the greater the spectral weight for
higher-harmonic phonon excitations (Fig. \ref{Fig3:Theory}(b)). In addition,
our calculation explicitly demonstrates that the difference between $g_{Cu}$
and $g_O$ can be resolved by comparing the RIXS spectrum taken at the UHB and
LEP resonances (Fig. \ref{Fig3:Theory}(c)). This site dependent information is
unique to RIXS and confirms the concepts illustrated in Fig. 1b.

\begin{figure} [t] \includegraphics [clip, width=3.5 in]{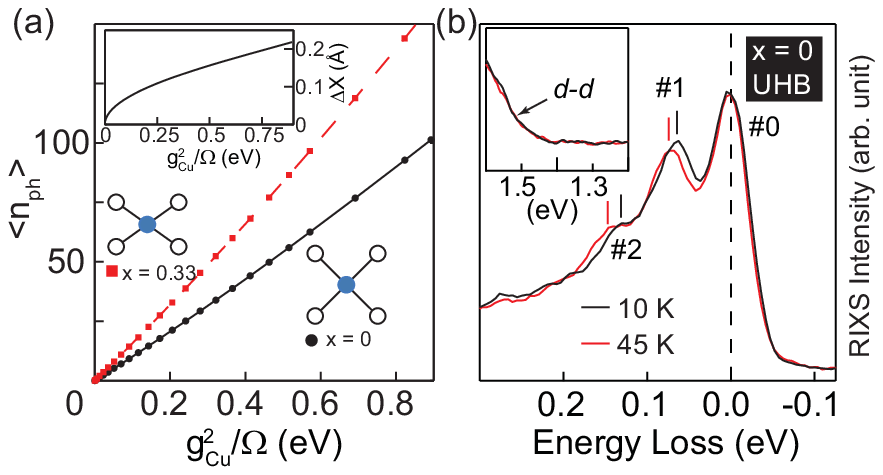}
    \caption{\label{Fig4:T_dep} (color online) (a) Calculated number of phonons
        entangled with the ground state wavefunction as a function of coupling
        strength for undoped and doped clusters. A larger number of phonon
        quanta in the ground state correspond to a larger lattice distortion.
        Please note that RIXS process excites additional phonons, leaving the
        system in an excited lattice state. The inset shows the estimated
        doping-induced changes in the lattice constant as a function of the
        mode coupling strength. (b) RIXS spectra taken at two temperatures
        across the AFM phase transition for the x = 0 compound. The inset shows
    the rising edge of the d-d excitations. The agreement of the elastic peak
and the d-d edge provide an internal energy reference, confirming the shift of
the phonon mode energy across the AFM transition.  } \end{figure}

With these insights, we can compare the electron-lattice coupling strength via
the spectral shape of the phonon excitations. First, as shown in Fig.
\ref{Fig3:Theory}(d), we found that the coupling strength is essentially doping
independent, even though a significant number of holes are doped into the
system. This demonstrates that screening remains ineffective upon doping, as
expected for a one dimensional system \cite{Brink00}. Second, the coupling at
the UHB resonance is found to be stronger than at the LEP resonance (Fig.
\ref{Fig3:Theory}(e)), which is caused by the difference between $g_{Cu}$ and
$g_O$. By fitting to our model, we estimate $g_{Cu} \sim 2g_O \sim 0.22$ eV, also in
agreement with our Madelung potential analysis (Supplementary Material).

A doping-independent electron-lattice coupling has an intriguing consequence to
the doping evolution of the intertwined electron-lattice ground state. As shown
in Fig. \ref{Fig4:T_dep}(a), the number of phonons intertwined in the ground
state is larger in the doped system than in the undoped system due to the
increased carrier concentration. This implies a larger lattice distortion along
the c-axis and results in a larger Cu-O-Cu angle in the doped system. Within
our model, we estimate a 0.16 Å contraction of O-O distance along c-axis,
corresponding to a 3.7\% increase of the Cu-O-Cu bond angle in the x = 0.33
doped cluster. The estimation is comparable to the reported doping induced
increase of the Cu-O-Cu angle ($\sim$2.5\%) \cite{Kudo05}. Since the size and sign
of the spin super-exchange coupling is sensitive to the Cu-O-Cu bond angle,
this result shows that e-ph coupling strength is an important underlying
mechanism to drive the lattice structure that hosts the spin dynamics of this
1D system.

Finally, as shown in Fig. \ref{Fig4:T_dep}(b), the phonon excitations exhibit a
softening of up to $\sim$10 meV when the system is cooled across the
antiferromagnetic transition temperature $T_N$ (30 K). Importantly, inelastic
neutron scattering measurements also have reported a spin excitation hardening
of approximately 1-2 meV when cooling to low temperatures \cite{Matsuda05},
demonstrating a cooperative interplay between the spin, charge and lattice
sectors. We also note that the coupling to the lattice is likely responsible
for the unusual spectral broadening for the spin and charge dynamics in
quasi-1D cuprates observed by angle-resolved photoemission spectroscopy
\cite{Kim06}, inelastic neutron scattering \cite{Matsuda01}, and RIXS
measurements \cite{Schlappa09,Schlappa12}. Our results emphasize that lattice
coupling in low dimensional materials needs to be considered together with the
spin and charge dynamics in order to obtain a holistic picture of the
underlying physics.

The authors thank J. M{\'a}lek, S.-L. Drechsler, and M. Berciu for useful
discussions.  This work was supported by the U. S. Department of Energy, Office
of Basic Energy Science, Division of Materials Science and Engineering under
the contractno. DE-AC02-76SF00515. S. J. acknowledges funding from FOM (The
Netherlands). This work was performed at the ADRESS beamline of the Swiss Light
Source using the SAXES instrument jointly built by Paul ScherrerInstitut,
Switzerland and Politecnico di Milano, Italy.


\begin{thebibliography}{99} \bibitem{Ashcroft76} N. W. Ashcroft and N. D.
        Mermin, Solid State Physics (Saunder College Publishing, 1976).

\bibitem{Tinkham96} M. Tinkham, Introduction to superconductivity (McGraw-Hill
    Book CO., 1996).

\bibitem{Gruner94} G. Gruner, Density Waves in Solids (ABP Perseus Publishing,
    1994).

\bibitem{Devereaux91} T. P. Devereaux and D. Belitz, Phys. Rev. B \textbf{43},
    3736 (1991).

\bibitem{Lanzara01} A. Lanzara \emph{et al.}, Nature \textbf{412}, 510 (2001).

\bibitem{Shen04} K. M. Shen, F. Ronning, D. H. Lu, W. S. Lee, N. J. C. Ingle,
    W. Meevasana, F. Baumberger, A. Damascelli, N. P. Armitage, L. L. Miller,
    Y. Kohsaka, M. Azuma, M. Takano, H. Takagi, and Z.-X. Shen, Phys. Rev.
    Lett. \textbf{93}, 267002 (2004).

\bibitem{Lee07} W. S. Lee, S. Johnston, T. P. Devereaux, and Z. X. Shen, Phys.
    Rev. B \textbf{75}, 195116 (2007).

\bibitem{Meevasana06} W. Meevasana, N. J. C. Ingle, D. H. Lu, J. R. Shi, F.
    Baumberger, K. M. Shen, W. S. Lee, T. Cuk, H. Eisaki, T. P. Devereaux, N.
    Nagaosa, J. Zaanen, and Z.-X. Shen, Phys. Rev. Lett. \textbf{96}, 157003
    (2006).

\bibitem{Johnston12} S. Johnston, I. M. Vishik, W. S. Lee, F. Schmitt, S.
    Uchida, K. Fujita, S. Ishida, N. Nagaosa, Z. X. Shen, and T. P. Devereaux,
    Phys. Rev. Lett. \textbf{108}, 166404 (2012).

\bibitem{Mannella05} N. Mannella, W. L. Yang, X. J. Zhou, H. Zheng, J. F.
    Mitchell, J. Zaanen, T. P. Devereaux, N. Nagaosa, Z. Hussain and Z.-X.
    Shen, Nature \textbf{438}, 474 (2005).

\bibitem{Vasiliu99} L. Vasiliu-Doloc, S. Rosenkranz, R. Osborn, S. K. Sinha, J.
    W. Lynn, J. Mesot, O. H. Seeck, G. Preosti, A. J. Fedro, and J. F.
    Mitchell, Phys. Rev. Lett. \textbf{83}, 4393 (1999).

\bibitem{Millis95} A. J. Millis, P. B. Littlewood, and B. I. Shraiman, Phys.
    Rev. Lett. \textbf{74}, 5144 (1995).

\bibitem{Muzuno98} Y. Mizuno, T. Tohyama, S. Maekawa, T. Osafune, N. Motoyama,
    H. Eisaki and S. Uchida, Phys. Rev. B \textbf{57}, 5326 (1998).

\bibitem{Voit95} J. Voit, Rep. Prog. Phys. \textbf{58}, 977 (1995).

\bibitem{Leib68} E. H. Lieb and F. Y. Wu, Phys. Rev. Lett \textbf{20}, 1445
    (1968).

\bibitem{Kim06} B. J. Kim, H. Koh, E. Rotenberg, S.-J. Oh, H. Eisaki, N.
    Motoyama, S. Uchida, T. Tohyama, S. Maekawa, Z.-X. Shen, and C. Kim, Nature
    Phys. \textbf{2}, 397 (2006).

\bibitem{Schlappa12} J. Schlappa, K. Wohlfeld, K. J. Zhou, M. Mourigal, M. W.
    Haverkort,	V. N. Strocov, L. Hozoi, C. Monney, S. Nishimoto, S. Singh, A.
    Revcolevschi, J.-S. Caux, L. Patthey, H. M. R{\o}nnow, J. van den Brink,
    and T. Schmitt, Nature \textbf{485}, 82 (2012).

\bibitem{Brink00} J. van den Brink and G. A. Sawatzky, Europhys. Lett.
    \textbf{50}, 447-453(2000).

\bibitem{hayashi98} A. Hayashi, B. Batlogg, and R. J. Cava, Phys. Rev. B
    \textbf{58}, 2678(1998).

\bibitem{Kudo05} K. Kudo, S. Kurogi, Y. Koike, T. Nishizaki, and N. Kobayashi,
    Phys. Rev. B \textbf{71}, 104413 (2005).

\bibitem{Matsuda05} M. Matsuda, K. Kakurai, S. Kurogi, K. Kudo, Y. Koike, H.
    Yamaguchi, T. Ito, and K. Oka, Phys. Rev. B \textbf{71}, 104414 (2005).

\bibitem{Goodenough55} J. B. Goodenough, Phys. Rev. \textbf{100}, 564 (1955).

\bibitem{Kanamori59} J. Kanamori, J. Phys. Chem. Solids \textbf{10}, 87 (1959).

\bibitem{Anderson63} P. W. Anderson, Solid State Phys. \textbf{14}, 99 (1963).

\bibitem{Matsuda01} M. Matsuda, H. Yamaguchi, T. Ito, C. H. Lee, K. Oka, Y.
    Mizuno, T. Tohyama, S. Maekawa, and K. Kakurai, Phys. Rev. B \textbf{63},
    180403(2001).

\bibitem{Okada03} K. Okada and A. Kotani, J. Phys. Soc. Japn. \textbf{72}, 797
    (2003).

\bibitem{Ament11} L. J. P. Ament, M. van Veenedaal, T. P. Devereaux, J. P.
    Hill, and J. van den Brink, Rev. Mod. Phys. \textbf{83}, 705 (2011).

\bibitem{Ament11_2} L. J. P. Ament, M. van Veenedaal, and J. van den Brink,Europhy.
    Lett. \textbf{95}, 27008 (2011).

\bibitem{Kuzian12} R. O. Kuzian, S. Nishimoto, S.-L. Drechsler,J. M{\'a}lek, S.
    Johnston, Jeroen van den Brink, M. Schmitt, H. Rosner, M. Matsuda, K. Oka,
    H. Yamaguchi, and T. Ito, Phys. Rev. Lett. \textbf{109}, 117207 (2012).

\bibitem{Schlappa09} J. Schlappa, T. Schmitt, F. Vernay, V. N. Strocov, V.
    Ilakovac, B. Thielemann, H. M. R{\o}nnow, S. Vanishri, A. Piazzalunga, X.
    Wang, L. Braicovich, G. Ghiringhelli, C. Marin, J. Mesot, B. Delley, and L.
    Patthey, Phys. Rev. Lett. \textbf{103}, 047401(2009).

\end{thebibliography}
\end{document}